\documentclass[%
 reprint,
 superscriptaddress,
 amsmath,amssymb,
 aps,
prb, citeautoscript,
]{revtex4-1}
\usepackage[version=3]{mhchem} 

\usepackage{graphicx}
\usepackage{dcolumn}
\usepackage{bm}
\usepackage{color, soul}
\usepackage{natbib}
\makeatletter
\newcommand*{\rom}[1]{\expandafter\@slowromancap\romannumeral #1@}
\usepackage[colorlinks, linkcolor=blue, citecolor=blue, urlcolor=blue]{hyper ref}
\makeatother
\usepackage{soul}

\begin{document}
\preprint{APS/123-QED}

\title{Machine learning insights into band gap properties in halide-based perovskites}

\author{Chadawan Khamdang}
\affiliation{Department of Electrical and Computer Engineering, State University of New York at Binghamton, Binghamton, NY 13902, USA}
\author{Mengen Wang}
\email{mewang@unc.edu}
\affiliation{Department of Physics and Astronomy, University of North Carolina at Chapel Hill, Chapel Hill, NC 27599, USA}

\begin{abstract}

Halide perovskites show great promise for applications in optoelectronic devices. 
The lead-free perovskites are attracting increasing interest due to their low toxicity and motivate the exploration of alternative compositions and structures, including A$_2$BX$_6$, A$_2$BB$^\prime$X$_6$, A$_3$B$_2$X$_9$, and A$_4$BX$_6$. 
Accurate predictions of a wide range of band gap energies are important for designing new materials.
It is also desired to generate a direct relationship between the structural and elemental descriptors and the band gap energies.
In this work, we develop machine learning models to predict band gap energies across various types of halide perovskites based on atomic and structural properties.
Algorithms including ensemble tree-based methods, random forest regression (RFR), gradient boosted regression trees (GBRT), and extreme gradient boosting (XGB) showed strong predictive accuracy.
We also analyzed feature importance to identify key descriptors, including B-site and X-site elemental properties, as well as the number of A- and B-site atoms, as primary factors influencing band gap energies.
These results improve our understanding of the ML models and provide guidance for designing new halide perovskite materials.

\end{abstract}

\maketitle
\section{\label{intro}Introduction}

Halide perovskites have attracted significant attention as promising materials for next-generation optoelectronic applications, including solar cells\cite{xu2023challenges, ramadan2023methylammonium, cui2023issues, tong2020wide}, light-emitting diodes (LEDs)\cite{quan2018perovskites, liu2021metal, fakharuddin2022perovskite, cheng2020multiple, zhang2021opportunities}, and photodetectors\cite{wang2021recent, zhou2023wide, miao2019recent, wang2022perovskite}. 
Their appealing combination of properties, such as tunable band gaps\cite{walsh2015principles, kong2016simultaneous, ouyang2019exploiting}, high optical absorption, and long carrier diffusion lengths\cite{shrestha2022long, akel2023relevance}, has driven significant interest in the field. 
In particular, perovskite solar cells (PSCs) based on lead halide ABX$_3$ structures have achieved remarkable power conversion efficiencies (PCEs), now surpassing 26.64\%.\cite{li2024high, wen2023heterojunction}

While lead halide perovskites have shown great promise, challenges related to the toxicity of Pb have motivated the search for alternative Pb-free perovskites.\cite{soltani2025metal, miah2025lead, zhang2022defect} 
Beyond the conventional ABX$_3$ framework, where A is a monovalent cation, B is a divalent metal, and X is a halide, several other families such as A$_2$BX$_6$, A$_2$BB$^\prime$X$_6$, A$_3$B$_2$X$_9$, and A$_4$BX$_6$ have garnered increasing interest.
These non-ABX$_3$ structures result from variations in the connectivity of [BX$_6$] octahedra and the presence of ordered cation or anion vacancies, which lead to diverse structures and a wide range of tunable band gaps.\cite{yi2016entropic, saliba2016cesium}
For example, A$_2$BX$_6$ is a vacancy-ordered double perovskite derived from ABX$_3$ by systematically removing half of the B-site cations.\cite{rahim2020geometric, cai2017computational} 
Similarly, A$_2$BB$^\prime$X$_6$ incorporates two distinct B-site cations arranged in a double perovskite configuration, which allows greater control over the electronic properties depending on the choice of B (B$^{+1}$) and B$^\prime$ (B$^{3+}$) elements.\cite{lei2021lead} 
These alternative structures offer promising directions toward more stable and lead-free perovskites.

Accurately predicting band gap energies across this diverse chemical and structural space is critical for identifying suitable candidates for device applications. 
However, experimental synthesis and characterization of all possible compositions are impractical because of the large number of possible material combinations. 
First principles calculations, such as density functional theory (DFT), provide a powerful alternative, but they can become computationally demanding, especially when hybrid functionals are required for accurate predictions.\cite{das2022density, janesko2021replacing} 
Machine learning (ML) has recently gained popularity in the materials science community as a data-driven approach to accelerate materials discovery and property prediction.
By learning patterns from existing data, ML models can rapidly predict key properties, such as band gap\cite{gou2025machine, li2021bandgap}, defect formation energy\cite{zhai2022predicting, khamdang2025defect}, and phase stability\cite{talapatra2021machine}, with reduced computational cost. 
Several studies have successfully applied ML to model the band gaps of perovskites and other semiconductors using features derived from atomic and structural descriptors. 
For instance, models trained on ABX$_3$\cite{gladkikh2020machine} and ABO$_3$ perovskites\cite{li2020progressive} have shown promising accuracy, and some efforts have extended to double perovskites\cite{lan2023comprehensive} and perovskite oxides\cite{talapatra2023band}.

Most of these studies have focused on a single perovskite family or a narrow compositional range, which limits their generalizability.
To date, few ML models have been developed to accommodate mixed A-site compositions or the broader chemical diversity observed in halide perovskites\cite{gou2025machine}. 
The application of ML across multiple halide perovskite families, such as A$_2$BX$_6$, A$_2$BB$^\prime$X$_6$, A$_3$B$_2$X$_9$, and A$_4$BX$_6$, remains largely unexplored.
In this work, we address this gap by developing machine learning regression models to predict the band gap energies of a wide range of halide perovskite structures, including ABX$_3$, A$_2$BX$_6$, A$_2$BB$^\prime$X$_6$, A$_3$B$_2$X$_9$, and A$_4$BX$_6$. 
Our dataset combines the band gap values from literature and new DFT data, particularly for A$_2$BX$_6$ compounds, where both cubic and rhombohedral structures are considered.
We construct input features based on elemental and structural properties, which are used to train four regression models: kernel ridge regression (KRR), random forest regression (RFR), gradient boosted regression trees (GBRT), and extreme gradient boosting (XGB).
In addition to evaluating predictive performance, we perform feature importance analysis to gain insights into the factors that influence band gap variation. 
This analysis not only improves the interpretability of the ML models but also provides guidance for future compositional design by identifying the key descriptors that control band gap behavior.

\section{\label{computational_details}Computational details}

\begin{figure}
    \includegraphics[width=90mm]{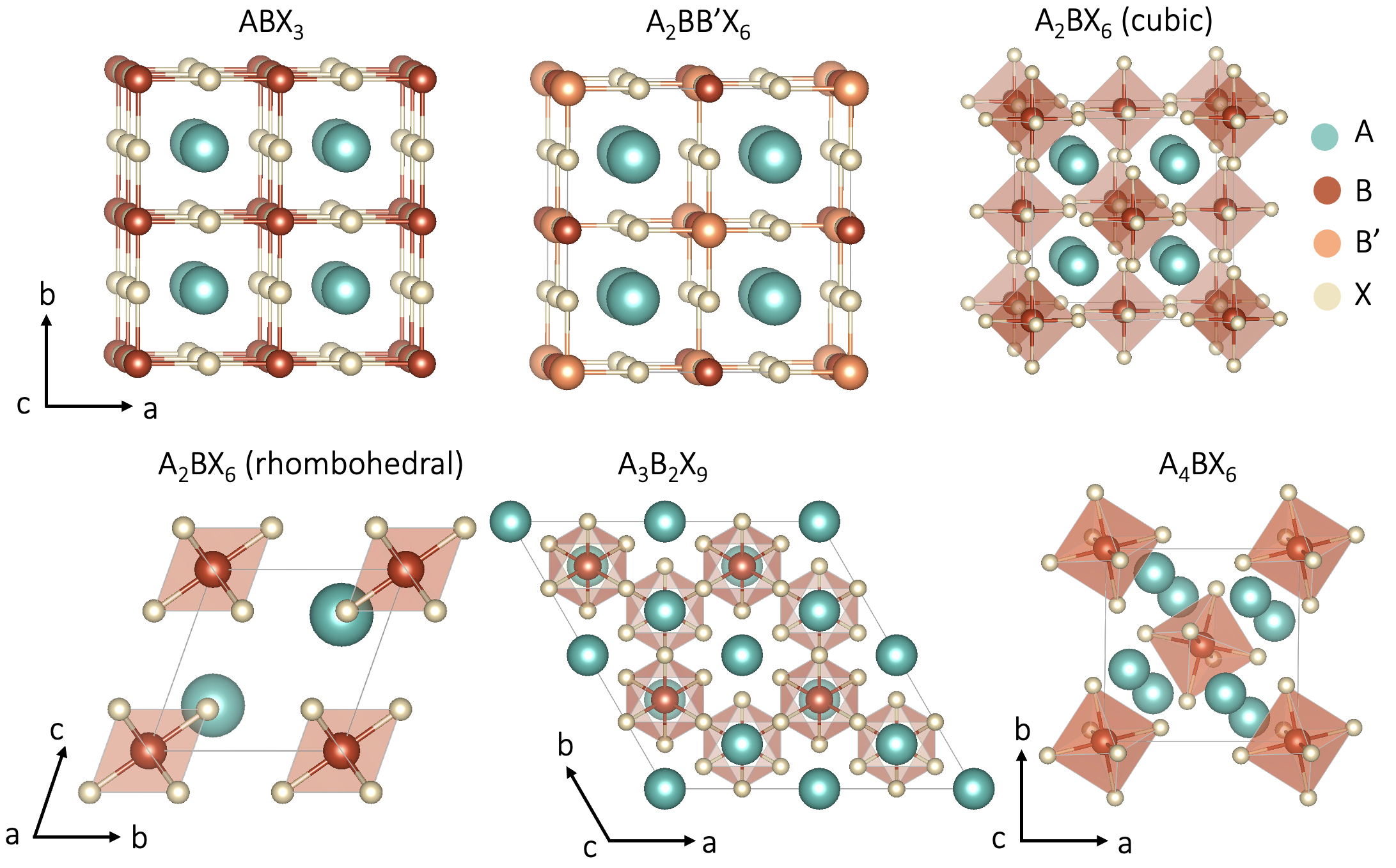}
    \caption{Crystal structures of the halide perovskite families investigated in this study, including cubic ABX$_3$, cubic A$_2$BX$_6$, face-centered cubic A$_2$BB$^\prime$X$_6$, rhombohedral A$_2$BX$_6$, hexagonal A$_3$B$_2$X$_9$, and trigonal A$_4$BX$_6$ structures.}
    \label{Fig1}
\end{figure}

To build the dataset, we collected data on halide perovskite structures, including ABX$_3$, A$_2$BX$_6$, A$_2$BB$^\prime$X$_6$, A$_3$B$_2$X$_9$, and A$_4$BX$_6$, from previous hybrid DFT calculations.\cite{park2021cost} 
The initial dataset consisted of 214 materials with non-zero band gaps.
Crystal symmetries were examined using the Pymatgen package.\cite{ong2013python} 
The crystal structures and symmetries of the halide perovskite families included in this study are shown in Fig. \ref{Fig1}.
These include cubic ABX$_3$, cubic A$_2$BX$_6$, face centered cubic A$_2$BB$^\prime$X$_6$, rhombohedral A$_2$BX$_6$, hexagonal A$_3$B$_2$X$_9$, and trigonal A$_4$BX$_6$ structures.

Band gap energies were calculated using density functional theory (DFT) as implemented in the Vienna Ab initio Simulation Package (VASP).\cite{29-vasp} 
Projector-augmented wave (PAW) pseudopotentials\cite{30-pseudoProjectoraugmented-wavemethod} were used with a plane-wave energy cutoff of 500 eV. 
The HSE06 hybrid functional\cite{31-Hybridfunctionals-based-on-screened-Coulomb-potential} was employed with a mixing parameter of 0.25. 
We considered A$_2$BX$_6$ structures with both a checkerboard-like pattern of vacancies (cubic phase) and isolated [MX$_6$] octahedra, which are referred to as quasi-zero-dimensional perovskites.\cite{maughan2019perspectives,jin2022unraveling,xu2018zero,liu2024pt,wang2024vacancy,zheng2022vacancy}
The cubic A$_2$BX$_6$ structures were derived from $2 \times 2 \times 2$ supercells of cubic ABX$_3$, which provide periodically ordered vacancies, while the isolated [MX$_6$] octahedra were modeled using a rhombohedral structure.
The Brillouin zone was sampled using a $\Gamma$-centered $2 \times 2 \times 2$ k-point mesh.
All structures were fully relaxed until the atomic forces were less than 0.01 eV/\AA.

\section{\label{results}Results and discussions}
\subsection{\label{effectonEg}Effect of crystal structure and composition on band gap energy
}

\begin{figure}
    \includegraphics[width=90mm]{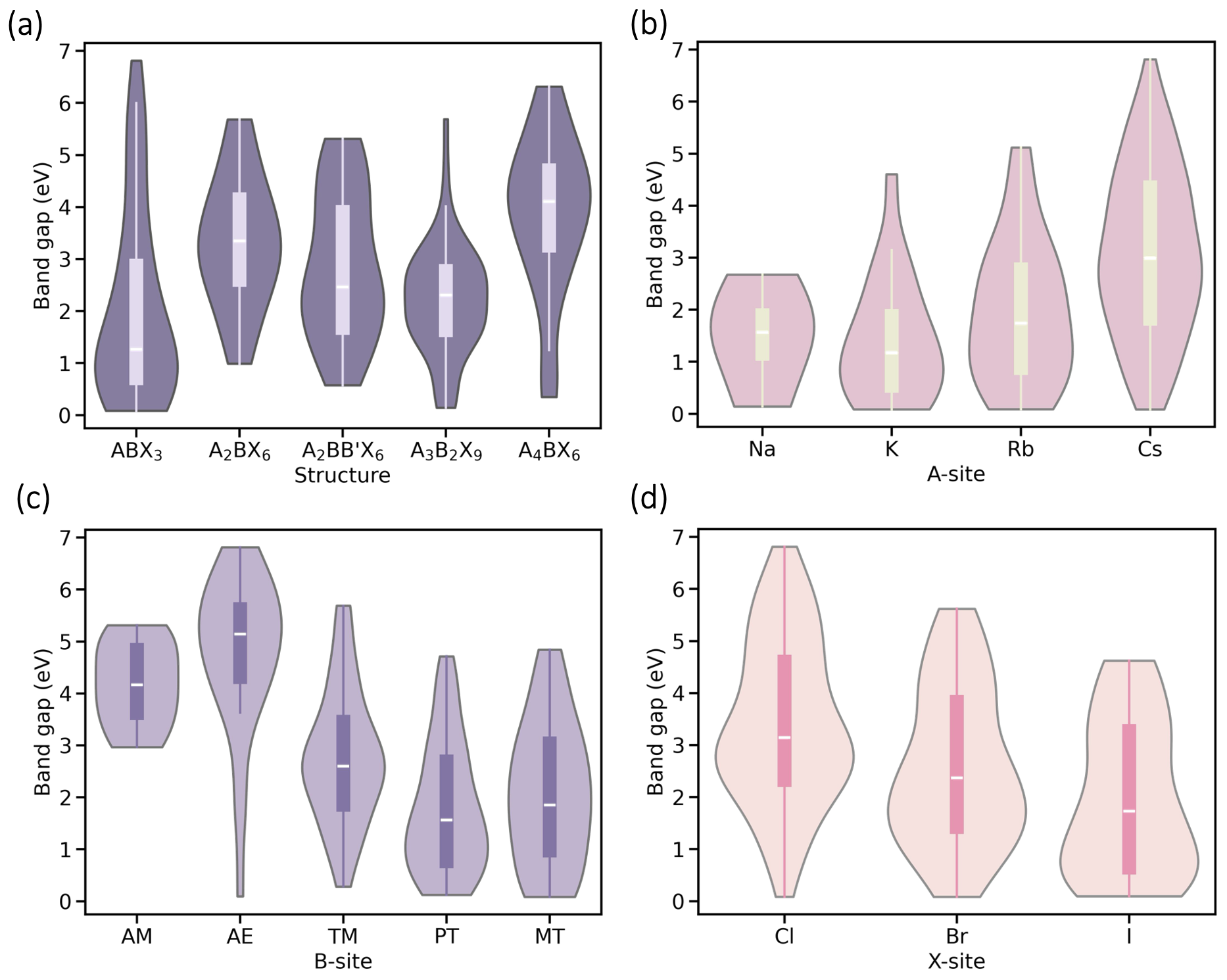}
    \caption{Violin plots showing the distribution of calculated band gap energies for halide perovskites across different categories: (a) structural families, (b) A-site cations, (c) B-site cations, and (d) X-site halides.
    Each violin plot represents the probability density of the data.
    The box represents the interquartile range (25\% to 75\%), the central line marks the median, and the vertical lines extend from the box, typically indicating the range of non-outlier data.
    The top and bottom ends of the plot represent the maximum and minimum band gap values.}
    \label{Fig2}
\end{figure}

To understand the influence of crystal structure and chemical composition on the electronic properties of halide perovskites, we first analyzed the distribution of band gap energies across various material families, as shown in the violin plots in Fig. \ref{Fig2}.

Fig. \ref{Fig2}(a) summarizes the band gap distributions in different crystal structures.
In cubic ABX$_3$ perovskites, the median band gap is 1.26 eV, which falls within the ideal range for single-junction solar cells (1.00–1.50 eV).\cite{shockley2018detailed}
This family also exhibits a broad band gap distribution, spanning from approximately 0.1 eV to over 7.0 eV, reflecting its wide-ranging compositional space and inherent tunability.
Cubic A$_2$BX$_6$ compounds, including Cs$_2$BX$_6$ (B = Ti, Sn, Te, Zr, and Hf; X = Cl, Br, and I), show band gaps ranging from 1.68 to 5.68 eV, which is consistent with previous DFT studies using PBEsol that reported values between 0.75 and 4.14 eV.\cite{zhang2025efficiently}
The median band gap of the cubic A$_2$BX$_6$ structure is 3.35 eV. 
A$_2$BB$^\prime$X$_6$ double perovskites exhibit band gaps ranging from 0.57 to 5.31 eV, making them suitable for a wide range of optoelectronic applications.\cite{park2021cost}
The A$_3$B$_2$X$_9$ family predominantly shows band gaps between 1.63 and 2.77 eV (interquartile range), representing a relatively narrow distribution that agrees well with previous reports (0.80–3.30 eV).\cite{luo2024computational}
Among all structural families, A$_4$BX$_6$ compounds exhibit the highest average band gap, with a median value of 4.10 eV.

Fig. \ref{Fig2}(b) categorizes the perovskites by A-site, and the band gap generally increases with the atomic number of the A-site element. For instance, the median band gap for K-containing compounds is 1.17 eV, rising to 1.74 eV for Rb and 2.99 eV for Cs. 
This trend aligns with previous reports, such as CsPbI$_3$ (1.59 eV) $>$ RbPbI$_3$ (1.57 eV) $>$ KPbI$_3$ (1.54 eV).\cite{ghosh2023lead} 
Larger A-site cations lead to less octahedral tilting, subtly modifying bond lengths and angles, and hence the band structure.\cite{hautzinger2024site}
Note that Na-containing compounds exhibit a higher average band gap compared to those containing K. 
This is mainly because Na appears exclusively in the A$_3$B$_2$X$_9$ family within our dataset, whereas other A–site cations, such as K, are present across multiple structural types.

Fig. \ref{Fig2}(c) shows the effects of the B-site. 
Alkaline metals (AM) and alkaline earth metals (AE) tend to exhibit higher median band gaps (AM: 4.17 eV, AE: 5.14 eV) compared to transition metals (TM: 2.60 eV), post-transition metals (PT: 1.56 eV), and metalloids (MT: 1.85 eV). 
This trend is primarily attributed to the electronegativity (EN) of the B-site elements. 
A lower EN leads to weaker hybridization between B and X orbitals, resulting in a wider band gap.\cite{walsh2015principles, yuan2015nature} 
For example, the band gap increases from CsSnCl$_3$ (1.08 eV) to CsCdCl$_3$ (3.14 eV) to CsMgCl$_3$ (5.55 eV), consistent with increasing EN: Mg (1.31) $<$ Cd (1.69) $<$ Sn (1.96). 
Similarly, Cs$_2$KInCl$_6$ (5.07 eV) exhibits a higher band gap than Cs$_2$NaInCl$_6$ (4.49 eV), in line with K (0.82) having a lower EN than Na (0.93). 
This pattern is also supported by previous findings, such as K$_2$InAlCl$_6$ (2.66 eV) $>$ K$_2$AgAlCl$_6$ (2.19 eV), reflecting the lower EN of In (1.78) compared to Ag (1.93).\cite{lan2023comprehensive} 
Post-transition metals also exhibit the narrowest band gap distributions, highlighting their promise for photovoltaic applications.

Considering the X-site effects shown in Fig. \ref{Fig2}(d), the band gap consistently decreases with increasing halogen atomic size. 
Median values are highest for Cl (3.14  eV), followed by Br (2.37 eV), and lowest for I (1.73 eV). 
This trend is attributed to the dominant contribution of halogen $p$-orbitals to the valence band maximum (VBM). 
The trend also aligns with the electronegativity order Cl $>$ Br $>$ I, where higher EN corresponds to lower-lying $p$-orbitals and a wider band gap. Our dataset reflects this trend clearly: CsPbCl$_3$ (1.78 eV) $>$ CsPbBr$_3$ (1.26 eV) $>$ CsPbI$_3$ (0.78 eV), consistent with previous studies.\cite{tao2019absolute}

\subsection{\label{features}Features for machine learning regression models}

We initially selected 44 features representing the atomic and structural properties of the constituent elements at the A, B, and X-sites. 
Atomic features included atomic weight (W), electronegativity (EN), electron affinity (EA), ionization energy (IE) (including 1st, 2nd, and 3rd), atomic radius (AR), covalent radius (CR), Shannon’s ionic radius (IR)\cite{shannon1976IR}, density (D), specific heat capacity (S), and heat of vaporization (HV). 
The structural features included the number of A, B, and X-site atoms (\# of A, \# of B, and \# of X), the Goldschmidt tolerance factor t(AR), and the octahedral factor $u$(IR), defined as\cite{goldschmidt1926gesetze, li2008formability}
\begin{equation}
    \mathrm{t(AR)} = \frac{r_\mathrm{A} + r_\mathrm{X}}{\sqrt{2}(r_\mathrm{B} + r_\mathrm{X})}
\end{equation}

\begin{equation}
    u\mathrm{(IR)} = \frac{r_\mathrm{B}}{r_\mathrm{X}}
\end{equation}
where $r_\mathrm{A}$, $r_\mathrm{B}$, and $r_\mathrm{X}$ are the atomic radii of the A, B, and X-site atoms, respectively. For the tolerance factor $\mathrm{t(AR)}$, $r$ values correspond to atomic radii, while for the octahedral factor $u(\mathrm{IR})$, $r$ values represent Shannon’s ionic radii.
For A$_2$BB$^\prime$X$_6$ compounds, B-site properties were averaged over B and B$^\prime$ atoms.

To reduce feature redundancy and select the most informative descriptors, we employed correlation analysis. 
We assessed linearity and monotonicity using Pearson\cite{cohen2009pearson}, Spearman\cite{spearman1987proof}, and Kendall's correlation coefficients\cite{kendall1938new}, as shown in Fig. S1. 
Among these, the Pearson correlation consistently provided the clearest and most direct measure of linear relationships with the band gap, compared to Spearman and Kendall. 
Exceptions were observed for the \# of A and the 3rd IE\_B feature, where Spearman correlation showed higher values (0.33 for \# of A and 0.30 for 3IE\_B) than Pearson correlation (0.31 and 0.21, respectively), while Kendall’s correlation exhibited lower coefficients for all features. However, Pearson generally yielded higher coefficients for the remaining features and was therefore considered the most suitable metric for our regression models.
If two features had a high absolute correlation ($|p| >$ 0.85), we retained the one with stronger correlation to the target property.
This filtering was applied independently to the A, B, and X-site features. 
As a result, 16 features were selected for machine learning model training.

\begin{figure}
    \includegraphics[width=90mm]{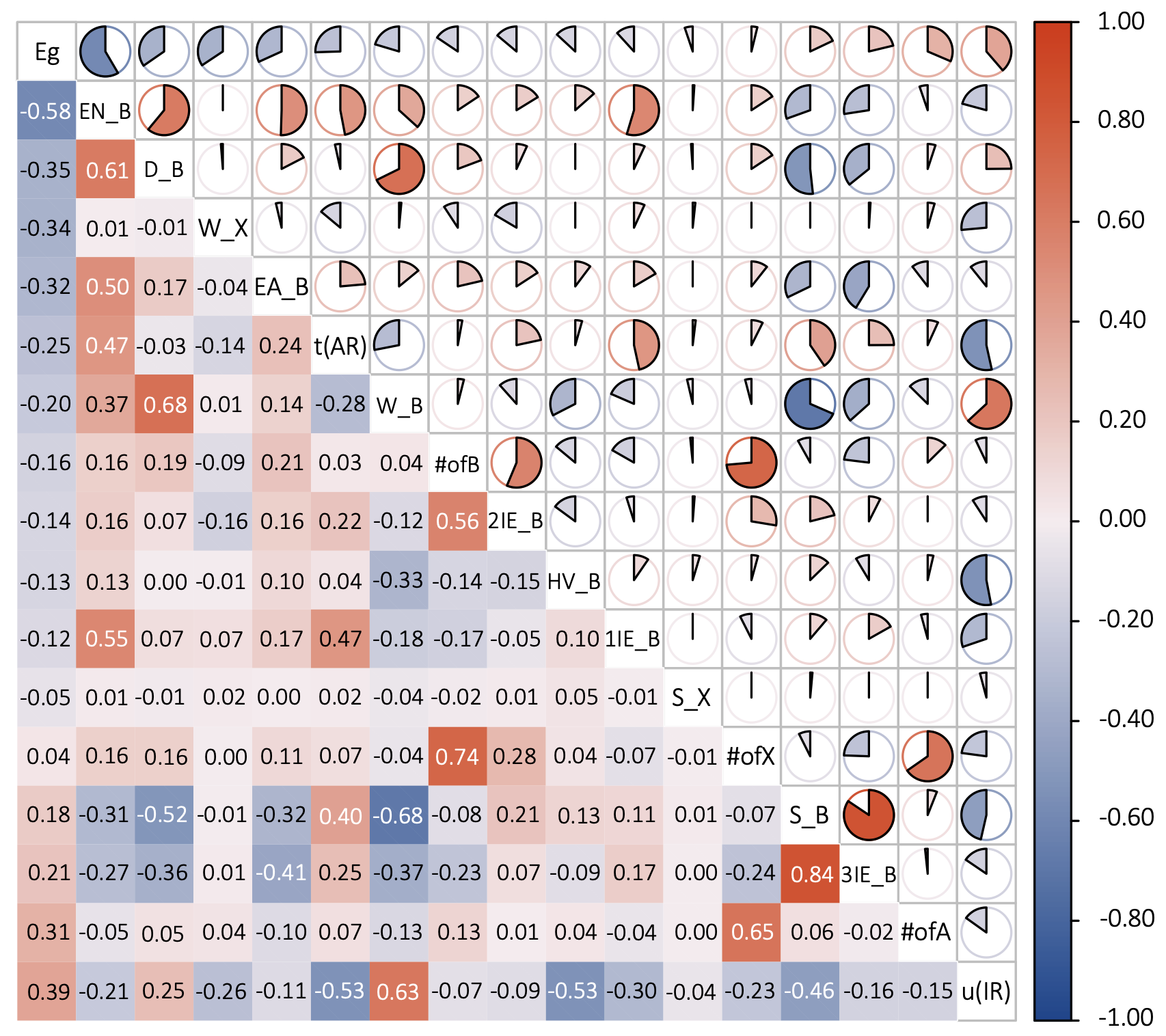}
    \caption{Pearson correlation matrix showing pairwise correlations between features and the band gap target property (E$_\mathrm{g}$). 
    The down-selected features are listed along the diagonal. 
    The upper and lower triangles present the same correlation data using two complementary visualization schemes. 
    Features are arranged from strong negative to strong positive correlation (left to right or top to bottom).}
    \label{Fig3}
\end{figure}

The correlations between the selected features and the band gap energy are shown in Fig. \ref{Fig3}.
Elemental properties of the B-site and X-site, as well as structural descriptors, show the strongest correlations with the band gap.
Elemental features of the B-site exhibit the strongest correlation with the band gap. 
In particular, EN\_B (electronegativity of the B-site) shows the strongest negative correlation ($p$ = -0.58), consistent with the trends discussed in Section \ref{effectonEg}, where lower B-site electronegativity is associated with wider band gaps. 
In addition, D\_B ($p$ = -0.35) and 3IE\_B ($p$ = 0.21) also show strong correlations with the band gap.
Among the X-site features, W\_X (atomic weight of the X-site) exhibits a moderate negative correlation ($p$ = -0.34) and was selected for its simplicity and consistency with related descriptors such as EN\_X and IE\_X. 
These features collectively capture the general trend of decreasing band gap with increasing halogen atomic number and size.
For structural features, the number of A-site atoms (\# of A) shows a positive correlation with the band gap ($p$ = 0.31), reflecting the trend of increasing band gap with higher A-site content. 
Both $t$(AR) and $u$(IR) also exhibit correlations of $p$ = -0.25 and $p$ = 0.39, respectively. 
In particular, $u$(IR) shows the strongest positive correlation with the band gap among the structural descriptors. 
This can be attributed to the B–X bonding.\cite{woodward1997octahedral,amat2014cation,filip2014steric}
These correlation trends are also consistent with those observed for rhombohedral A$_2$BX$_6$ compounds, as shown in Fig. S1.

We then trained four machine learning (ML) models, including kernel ridge regression (KRR), random forest regression (RFR), gradient-boosted regression trees (GBRT), and extreme gradient boosting (XGB) as implemented in the \texttt{scikit-learn} package\cite{pedregosa2011scikit}, to predict the band gap energy using the selected features. 
These models were chosen to represent a diverse set of regression algorithms, balancing interpretability with the ability to capture complex, non-linear relationships inherent in materials data. 
This selection was also informed by findings from previous studies, which indicate that linear regression models often fail to capture the nonlinear relationships inherent in materials data.\cite{khamdang2025defect, lan2023comprehensive}

Model evaluation was performed using Leave-One-Out Cross Validation (LOOCV). 
Hyperparameters were optimized using a grid search strategy, with five-fold cross-validation used within the grid search to prevent overfitting.\cite{25-predictml2impuritiesinhalideperovskites, 26-predictml3predictabo3}
Model performance was evaluated using three metrics, including coefficient of determination ($R^2$), root mean square error (RMSE), and mean absolute error (MAE). 
$R^2$ quantifies the proportion of variance in the target explained by the model, with values closer to 1 indicating better performance. 
RMSE measures the standard deviation of the prediction errors, providing a sense of the typical error magnitude, while MAE represents the average absolute error, offering a robust measure less sensitive to outliers. The formulas for these metrics are
\begin{equation}
    R^2 = 1 - \frac{\sum_{i=1}^{n} (y_i - \hat{y}_i)^2}{\sum_{i=1}^{n} (y_i - \bar{y})^2}
\end{equation}

\begin{equation}
    \text{RMSE} = \sqrt{\frac{1}{n} \sum_{i=1}^{n} (y_i - \hat{y}_i)^2}
\end{equation}

\begin{equation}
    \text{MAE} = \frac{1}{n} \sum_{i=1}^{n} |y_i - \hat{y}_i|
\end{equation}
where \( y_i \) is the true value, \( \hat{y}_i \) is the predicted value, and \( n \) denotes the number of samples in the dataset. 
The term \( \bar{y} \) is the average of the actual values across all samples in the dataset.

Additionally, we evaluated feature importance derived from the trained models and systematically compared these results with the Pearson correlation coefficients to extract deeper physical insights into the band gap driving forces and improve the overall interpretability of our models.

\begin{figure*}
    \includegraphics[width=7in]{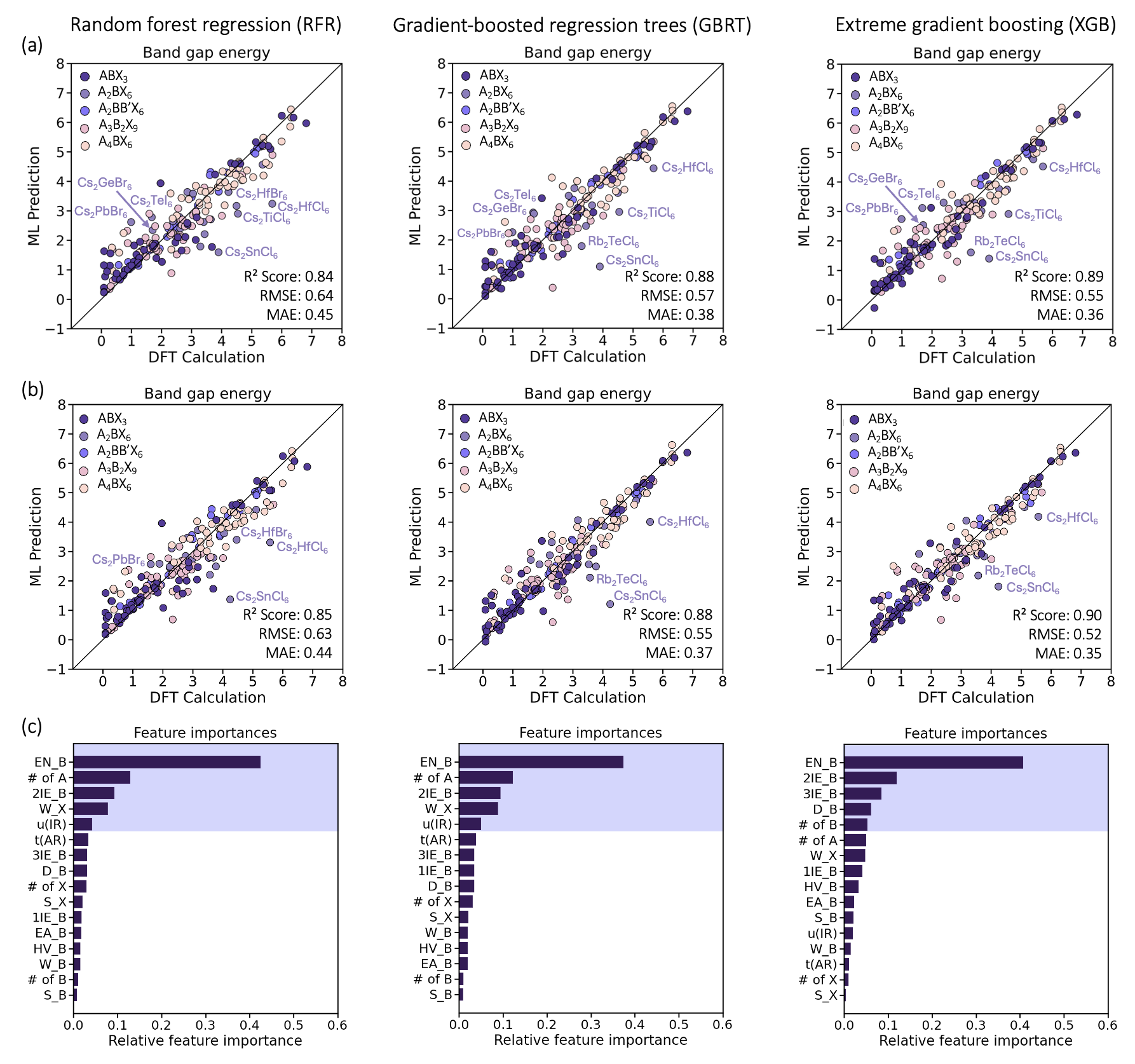} 
    \caption{Parity plots from random forest regression (RFR), extreme gradient boosting (XGB), and gradient boosted regression trees (GBRT) for (a) reference data of A$_2$BX$_6$ in the cubic phase and (b) DFT calculated data of A$_2$BX$_6$ in the rhombohedral phase. (c) Feature importance for the three regression models. The colors in the parity plots distinguish between different halide perovskite types, and the shaded purple region in the feature importance plots highlights the top five most important features.} 
    \label{Fig4} 
\end{figure*}

\subsection{\label{ML}Training machine learning models}

Kernel ridge regression (KRR) is a nonlinear extension of ridge regression that uses kernel functions to capture complex patterns in the data.\cite{vovk2013kernel} 
Both linear and radial basis function (RBF) kernels were evaluated to determine the best fit for our dataset.\cite{duvenaud2014automatic}
Random forest regression (RFR) is an ensemble learning method that constructs multiple decision trees and aggregates their predictions to improve accuracy and reduce overfitting.\cite{rfr}
Model performance was optimized by tuning key hyperparameters, including the number of estimators (trees), maximum tree depth, number of leaf nodes, and the maximum number of features considered at each split.

Gradient-boosted regression trees (GBRT) are another tree-based ensemble method that builds trees sequentially, with each tree learning to correct the errors of its predecessors. 
Key hyperparameters include the learning rate (which controls the contribution of each tree), the number of trees, and their maximum depth.\cite{natekin2013gradient}
Extreme gradient boosting (XGB) is an optimized implementation of gradient boosting known for its high performance and computational efficiency.
XGB incorporates regularization techniques (L1 and L2), efficient tree pruning, and parallel processing to enhance generalization and mitigate overfitting.\cite{chen2015xgboost}

Among the evaluated regression models, RFR, GBRT, and XGB demonstrate stronger predictive performance. 
The corresponding parity plots are shown in Fig. \ref{Fig4}(a).
The RFR model achieves $R^2$/RMSE/MAE of 0.84/0.64/0.45 eV. Further improvement is observed for GBRT and XGB, with $R^2$/RMSE/MAE values of 0.88/0.57/0.38 eV and 0.89/0.55/0.36 eV, respectively. 
The parity plots for the KRR model across all datasets are shown in Fig. S2. 
The KRR model yields an $R^2$ of 0.72, an RMSE of 0.86 eV, and an MAE of 0.67 eV, indicating relatively lower predictive accuracy compared to the other models.

For the three models trained on the reference dataset, A$_2$BX$_6$ compounds are observed to deviate from the diagonal parity line. 
To investigate this behavior, we generated A$_2$BX$_6$ structures using two approaches and calculated the band gap values using the HSE06 functional as described in the computational details.
First, cubic A$_2$BX$_6$ structures were generated from a $2 \times 2 \times 2$ cubic ABX$_3$ supercell by removing selected B-site cations to create an ordered vacancy arrangement. 
Second, rhombohedral A$_2$BX$_6$ structures were obtained directly from the rhombohedral crystal phase, in which the vacancy ordering is already incorporated into the crystallographic framework.
In the original dataset, all A$_2$BX$_6$ compounds were reported in the cubic Fm$\bar{3}$m phase. The energies per atom, together with the calculated and experimental band gap energies of cubic and rhombohedral A$_2$BX$_6$, are summarized in Table S1.

Our DFT calculations for A$_2$BX$_6$ in the cubic phase (Fig. S3) and the rhombohedral phase (Fig. \ref{Fig4}(b)) result in slightly improved predictive performance for all three models (RFR, GBRT, and XGB). 
This improvement is due to the existence of multiple structural representations of A$_2$BX$_6$, resulting in a distribution of band gap values. 
Cs$_2$TeI$_6$ and Cs$_2$GeBr$_6$ have smaller band gap energies in the reference dataset (1.68 eV and 1.71 eV), whereas our calculated band gaps are higher, 2.17/2.20 eV for the cubic phase and 2.29/2.23 eV for the rhombohedral phase, bringing these data points closer to the diagonal parity line. 
On the other hand, Cs$_2$TiCl$_6$ has a larger band gap of 4.45 eV in the original dataset, while our calculations yield smaller band gaps for both the cubic and rhombohedral phases and closer to experimental values (Table S1).
Overall, model performance improves progressively from KRR to RFR, GBRT, and XGB, and the incorporation of our calculated cubic and rhombohedral A$_2$BX$_6$ band gaps further enhances predictive performance, particularly for the rhombohedral phase.

During the training of the RFR, GBRT, and XGB models, we evaluated feature importance for band gap prediction. 
The top five most important features across these models are shown in Fig. \ref{Fig4}(c).
For both RFR and GBRT, the top five features are EN\_B, \# of A, 2IE\_B, W\_X, and $u$(IR).
These features are consistent with the highly correlated descriptors identified earlier, such as EN\_B ($p$ = -0.58) and W\_X ($p$ = -0.34), which show the strongest negative correlations with band gap, while \# of A ($p$ = 0.31) and $u$(IR) ($p$ = 0.39) are the most positively correlated. 
These findings align with previous studies highlighting the role of EN\_B in A$_2$BB$^\prime$X$_6$ compounds.\cite{lan2023comprehensive} 
Although 2IE\_B has only a moderate correlation ($p$ = -0.14), it is still one of the top-ranked features, showing that moderate correlations can still be important for prediction.
In contrast, the top five features for XGB are EN\_B, 2IE\_B, 3IE\_B, D\_B, and \# of A. 
Importantly, all top features in XGB are B-site properties. Among these, EN\_B ($p$ = -0.58), D\_B ($p$ = -0.35), and 3IE\_B ($p$ = 0.21) exhibit strong correlations, while 2IE\_B ($p$ = -0.14) and \# of B ($p$ = -0.16) show moderate correlations.
We note that the feature importance in Fig. \ref{Fig4}(c) is derived from the original dataset, and the top five important features remain the same for the rhombohedral structures (Fig. S4) and are similar to those from the cubic A$_2$BX$_6$ calculations (Fig. S3).
Overall, the analysis suggests that descriptors from the B and X-site elements, as well as structural properties such as the number of A and B-site atoms and $u$(IR), play a crucial role in determining band gap energies. 
Although the performances of RFR, GBRT, and XGB are similar, the top features consistently correspond to those with higher correlations EN\_B, D\_B, W\_X, 3IE\_B, \# of A, and $u$(IR). 
These results confirm that the top features are consistent with the strongly correlated descriptors and highlight the significance of B-site elemental properties in predicting band gap energies across various halide perovskite families, including ABX$_3$, A$_2$BX$_6$, A$_2$BB$^\prime$X$_6$, A$_3$B$_2$X$_9$, and A$_4$BX$_6$.

\section{\label{conclution}Conclusion}

In conclusion, we present a machine learning approach for predicting band gap energies across various halide perovskite structures, including ABX$_3$, A$_2$BX$_6$, A$_2$BB$^\prime$X$_6$, A$_3$B$_2$X$_9$, and A$_4$BX$_6$. 
Analysis of band gap distributions across different structural families and elemental sites provided key insights into the factors governing electronic properties. 
We observed that the number of A-site and B-site atoms influences the band gap, with B-site elements from the alkaline and alkaline earth metals exhibiting significantly higher band gaps compared to other groups, highlighting the importance of B-site properties such as electronegativity and ionization energy.
Additionally, the X-site trend shows a consistent decrease in band gap with increasing halogen atomic size.

To build predictive models, we selected 16 atomic and structural features using Pearson correlation, effectively capturing linear relationships with band gap energy. 
The ML models considered include kernel ridge regression (KRR), random forest regression (RFR), gradient-boosted regression trees (GBRT), and extreme gradient boosting (XGB). 
Among these, the ensemble tree-based models (RFR, GBRT, XGB) consistently outperformed KRR, achieving higher $R^2$ scores and lower RMSE and MAE values. 
Beyond prediction, feature importance analysis identified key elemental and structural descriptors for band gap prediction, including EN\_B, D\_B, W\_X, 3IE\_B, 2IE\_B, \# of B, \# of A, and $u$(IR).
The strong alignment between these top features and their Pearson correlations, combined with the ability of tree-based models to capture nonlinear effects, provides a predictive framework for estimating band gap energies in halide perovskite families.

\section*{Data and code availability}
Supporting information and data will be available after publication.

\section*{Acknowledgments}
This work was supported by the new faculty start-up grant from UNC-Chapel Hill. This work used Bridges-2 at Pittsburgh Supercomputing Center through allocation MAT230043 from the Advanced Cyberinfrastructure Coordination Ecosystem: Services \& Support (ACCESS) program, which is supported by National Science Foundation grants \#2138259, \#2138286, \#2138307, \#2137603, and \#2138296.

\bibliography{reference}

\end{document}